\documentclass{article}
\usepackage{spconf,amsmath,graphicx}
\usepackage{etoolbox}

\usepackage{times}
\usepackage{epsfig}
\usepackage{graphicx}
\usepackage{amsmath}
\usepackage{amssymb}
\usepackage{multirow}

\usepackage{microtype}
\usepackage{graphicx}
\usepackage{subfigure}
\usepackage{booktabs} 
\usepackage{amsthm,amsmath,amssymb}
\usepackage{mathrsfs}
\usepackage{bbm}
\usepackage{stfloats}
\usepackage{graphicx} 
\usepackage{float} 
\usepackage{subfigure} 
\usepackage{hyperref}
\usepackage{stfloats}
\usepackage{color}
\usepackage{algorithm}
\usepackage{algorithmicx}
\usepackage{algpseudocode}
\usepackage{textcomp}
\usepackage{amsmath}
\usepackage{hyperref}

\title{DEEP MANIFOLD TRANSFORMATION FOR PROTEIN REPRESENTATION LEARNING}
\name{Bozhen Hu$^{1, 2}$, Zelin Zang$^{2}$, Cheng Tan$^{2}$, Stan Z. Li$^{2 \star}$\thanks{`$\star$' denotes the corresponding author.}}
\address{$^{1}$ Zhejiang University, Hangzhou, 310058, China\\
$^{2}$AI Division, School of Engineering, Westlake University, Hangzhou, 310030, China}
\begin{document}
%
   \maketitle
\begin{abstract}
Protein representation learning is critical in various tasks in biology, such as drug design and protein structure or function prediction, which has primarily benefited from protein language models and graph neural networks. These models can capture intrinsic patterns from protein sequences and structures through masking and task-related losses. However, the learned protein representations are usually not well optimized, leading to performance degradation due to limited data, difficulty adapting to new tasks, etc. To address this, we propose a new \underline{d}eep \underline{m}anifold \underline{t}ransformation approach for universal \underline{p}rotein \underline{r}epresentation \underline{l}earning (DMTPRL). It employs manifold learning strategies to improve the quality and adaptability of the learned embeddings. Specifically, we apply a novel manifold learning loss during training based on the graph inter-node similarity. Our proposed DMTPRL method outperforms state-of-the-art baselines on diverse downstream tasks across popular datasets. This validates our approach for learning universal and robust protein representations. We promise to release the code after acceptance. 
\end{abstract}
\section{Introduction}
\label{sec:intro}

Proteins play essential roles in a variety of applications in biology, such as catalyzing biochemical reactions, immune function, and molecular recognition~\cite{zhang2022protein}, etc. Protein sequences are built from 20 different basic amino acids, which can be folded into structures, determining their functions~\cite{hu2022protein}. Learning representations of proteins with sequences and structures is crucial for a wide range of tasks~\cite{BowenJing2020LearningFP}. There exist similarities between human languages and protein sequences. For example, there are four different levels of protein structures: the amino acid chains, the local folded structures, the three-dimensional (3D) structure, and the complex that can analogy to letters, words, sentences, and texts in human languages. Therefore, graph neural networks (GNNs) and language models (LMs) are usually developed to learn from protein data~\cite{zhang2022protein,tan2022generative,tan2023global}. Existing protein representation learning methods are mainly categorized into three types: protein LMs for sequences, structure models for geometry, and hybrid methods for both. For instance, ProtTrans~\cite{AhmedElnaggar2021ProtTransTC} and ESM-1b~\cite{AlexanderRives2019BiologicalSA} utilize transformers to learn different ranges of residue dependencies. CDConv~\cite{fancontinuous} proposes continuous-discrete convolution to model geometry and sequences. Moreover, several methods model multiple structure levels or extra sources of information jointly~\cite{zhang2022protein}.

However, most of the protein representations learned from sequences or structures tend to be unconstrained by vanilla deep-learning techniques without specific operations based on protein-related knowledge. There is no doubt that proteins have their own principles, e.g., the hydrophilicity and hydrophobicity of amino acids~\cite{xu2022opus}, the principle of minimal frustration, and the "folding funnel" landscapes of proteins~\cite{leopold1992protein}, etc. Particularly, OmegaFold~\cite{RuidongWu2022HighresolutionDN} indicates that protein presentations without triangle attention modules or constraints may not satisfy the triangle inequality of amino-acid distances, which hinders the predictions on structure-related tasks. Biological tasks tend to have a limited number of samples, which hinders the ability to obtain universal protein representations.

Manifold learning tackles manifold representation problems under unsupervised conditions based on solid theoretical foundations. For example, MGAE~\cite{ChunWang2017MGAEMG} develops the auto-encoder for the graph area to embed node features and an adjacency matrix. Particularly, the manifold learning model can be seen as a general form of contrastive learning methods~\cite{zang2022dlme}, which have exhibited excellent performance on different tasks with the aid of data augmentation strategies. Inspired by these, we can develop deep manifold learning methods to preserve structural information to get more general and comprehensive protein representations; the manifold loss serves as guidance along with task losses~\cite{zang2022dlme}. In this paper, we propose a novel deep manifold framework, DMTPRL, to solve the above-mentioned problems and improve the quality of the representations. Firstly, we conduct data augmentations fitted with protein sequences or structures as input, and the commonly used protein sequential and structural features are calculated. Secondly, structural information or spatial relationships are obtained for subsequent operations. This process is achieved by obtaining the graph similarities collected from the relative positions of amino acids and node features. A long-tailed $t$-distribution is designed as a kernel function to fit the neighborhoods between nodes and transform the distances into similarities. Finally, a manifold learning loss is proposed to preserve the node similarities in the latent space, which can constrain original embeddings by preserving the distance-based spatial relationships. Thus, our contributions can be summarized as follows:
\begin{itemize}
\item We are the first to introduce manifold transformations in protein science to fit with the triangle distance properties of residues properties.
\item This proposed framework can be used for protein sequences or structures; the topological and geometric features are preserved by a specific manifold learning loss through considering the node-to-node similarity.
\item Achieve state-of-the-art performance on different tasks on the benchmark through extensive experiments.
\end{itemize}

\section{Method}
\label{sec:Method}

\subsection{Problem Statement}
We represent a protein graph as $G=(\mathcal{V}, \mathcal{E}, X, E)$, where $\mathcal{V}=\{v_i\}_{i=1, \ldots, n}$ and $\mathcal{E}=\left\{\varepsilon_{i j}\right\}_{i, j=1, \ldots, n}$ denote the vertex and edge sets with $n$ residues, respectively. Each amino acid has a 3D coordinate of $\mathrm{C}_\alpha$ that specifies its spatial position, thus, we use $\mathcal{P}=\{P_i\}_{i=1, \ldots, n}$ to represent the coordinates for nodes, where $P_i \in \mathbb{R}^{3 \times 1}$. The node and edge feature matrices are $X=[\boldsymbol{x}_i]_{i=1, \ldots, n}$ and $E=[\boldsymbol{e}_{ij}]_{i, j=1, \ldots, n}$, where $\boldsymbol{x}_i \in \mathbb{R}^{d_v}$ is the feature vector of node $v_i$, and $\boldsymbol{e}_{i j} \in \mathbb{R}^{d_\varepsilon}$ is the feature vector of edge $\varepsilon_{ij}$, $d_v$ and $d_\varepsilon$ mean the feature dimensions. The position or edge feature matrices $\mathcal{P}, E$, may not exist for proteins with only sequences as the input. 

\subsection{Protein Graph Construction}
We treat each amino acid as a graph node, using the concatenation of the one-hot encoding of residue types and the physicochemical properties of each residue, i.e., a steric parameter, hydrophobicity, volume, polarizability, isoelectric point, helix probability, and sheet probability~\cite{xu2022opus}. These physicochemical descriptors capture the inherent properties of amino acids that influence their patterns and roles in protein structure, folding, stability, and interactions. These properties provide insights that are hard to obtain from sequence or structure alone. 

Data augmentation schemes have been widely used in self-supervised learning to solve problems in computer vision and natural language processing. If the protein structures are known, we do data augmentations on positions and features to improve model performance. Otherwise, the augmentations are only conducted on the input feature matrix $X$. In biology, a linear combination of original data with Gaussian noise is a simple but effective way:
\begin{equation}
(P_i,\boldsymbol{x}_i)\gets (P_i,\boldsymbol{x}_i)+\Theta , \Theta  \sim (\mu_k,\sigma _k^2)
\label{(3)}
\end{equation}
where $\mu_k$ and $\sigma_k$ are selected as the random noise's mean (expectation) and standard deviation. 

The radius graph can be obtained by calculating the distance $d_{ij}=\left \| P_i-P_j \right \| $ between two nodes $v_i$ and $v_j$ when the positions are given, where $\left \| \cdot  \right \| $ denotes the $l^2$-norm. The edge $\varepsilon_{i j}$ exists when $d_{ij} < r$, where $r$ is a predefined threshold.

Considering the equivalent properties of protein structures, we construct the local coordinate system~\cite{ingraham2019generative} $\boldsymbol{Q}_i$ based on equivalent geometric relationships.
\begin{equation}
    \boldsymbol{Q}_i=[\boldsymbol{b_i} \quad \boldsymbol{n_i} \quad  \boldsymbol{b_i}\times \boldsymbol{n_i}]
\label{(4)}   
\end{equation}
where $\boldsymbol{u}_i=\frac{{P}_{i}-{P}_{i-1}}{\left\|{P}_{i}-{P}_{i-1}\right\|}, \boldsymbol{b_i}=\frac{\boldsymbol{u}_i-\boldsymbol{u}_{i+1}}{\left\|\boldsymbol{u}_i-\boldsymbol{u}_{i+1}\right\|}, \boldsymbol{n}_i=\frac{\boldsymbol{u}_i \times \boldsymbol{u}_{i+1}}{\left\|\boldsymbol{u}_i \times \boldsymbol{u}_{i+1}\right\|}$. 
Let $\mathcal{F}(\cdot)$ be the geometric transformation, then
\begin{equation}
    \mathcal{F}(G)_{ij} = (d_{ij},\boldsymbol{Q}_i^T\cdot \frac{{P}_{i}-{P}_{j}}{d_{ij}}, \boldsymbol{Q}_i^T\cdot \boldsymbol{Q}_j) 
\label{(5)}
\end{equation}
$\mathcal{F}(G)_{ij}$ is the complete geometric representation for protein 3D structures. We obtain the edge features by concatenating the relative sequential distances between nodes $v_i$ and $v_j$:
\begin{equation}
    \boldsymbol{e}_{ij} = \mathrm{Cat}(\mathcal{F}(G)_{ij}, \left |i-j  \right | )  
\label{(6)}
\end{equation}
where $\mathrm{Cat}(\cdot)$ denotes the concatenation operation, and $\left | \cdot  \right | $ means $l^1$-norm. 

\subsection{Deep Manifold Transformation}
\begin{figure}[htbp!]
\centering
\includegraphics[width=0.95\columnwidth]{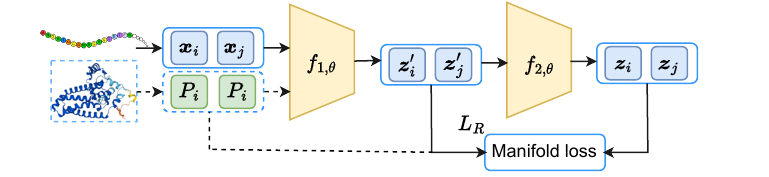} 
\caption{A pipeline to illustrate the process of deep manifold transformation, $\theta$ is the network parameter.}
\vspace{-1em}
\label{fig1}
\end{figure}

As shown in Figure~\ref{fig1}, the framework is mainly divided into two parts, $f_{1, \theta}, f_{2, \theta}$. There are several ($L_F$) fully-connected (FC) layers in $f_{1, \theta}$, which are designed to fit the augmented data and map these data into a latent space that better guarantees local connectivity and smoothness. Then $f_{2, \theta}$ is a graph convolutional network~\cite{kipf2016semi} with $L_M$ layers.

We aim to preserve the topological structure of latent embeddings by maintaining the inter-node similarity between two latent spaces, which are the output spaces of $f_{1, \theta}, f_{2, \theta}$. We build two graphs $G_{Z',\mathcal{P}}=(\mathcal{V}, \mathcal{E}, Z', E_{f_1})$ and $G_{Z}=(\mathcal{V}, \mathcal{E}, Z, E_{f_2})$ in the two latent spaces, respectively. Obviously, the two graphs have different node ($Z'$ and $Z$) and edge ($E_{f_1}$ and $E_{f_2}$) features. Define the graph distance matrix for proteins. $D^{G_{Z',\mathcal{P}}}=\left \{ d^{G_{Z',\mathcal{P}}}_{ij}|i,j=1,2, \cdots,n  \right \} $, when $\mathcal{P}$ is accessible, we have:
\begin{equation}
d_{ij}^{G_{Z',\mathcal{P}}}=\left\{\begin{matrix} \left \| P_i-P_j \right \|
 & \mathrm{if} \  \varepsilon_{i j} \in E \\ \Lambda 
  & \mathrm{otherwise}
\end{matrix}\right.
\label{(7)}
\end{equation}
Where $\Lambda$ is a large constant number. If $\mathcal{P}$ does not exist, we have $d_{ij}^{G_{Z',\mathcal{P}}}=\left \| \boldsymbol{z}'_i-\boldsymbol{z}'_j \right \|$. For a given graph $G_{Z}=(\mathcal{V}, \mathcal{E}, Z, E_{f_2})$, $d_{ij}^{G_{Z}}=\left \| \boldsymbol{z}_i-\boldsymbol{z}_j \right \|$. Thus, we can obtain the distance matrices $D^{G_{Z',\mathcal{P}}}$ and $D^{G_{Z}}$ in the two latent spaces.

Considering the class imbalance that may exist in protein data, here we adopt a long-tailed $t$-distribution to convert the distance matrix into a similarity matrix, while t-SNE~\cite{van2008visualizing} and UMAP~\cite{mcinnes2018umap} have adopted the normalized Gaussian and Cauchy functions and the fitted polynomial function as the transform function. $t$-distribution is more robust to outliers in the distance data than the Gaussian or Cauchy distributions, which provide a smooth monotonic conversion from distances to similarities. Moreover, the degree of freedom, $\nu$ in $t$-distribution, can be used as a tool to prevent the training from converging into bad local minima and control the separation margin between different
manifolds in experiments. Using $d^G_{ij}$ to represent $d_{ij}^{G_{Z',\mathcal{P}}}$ or $d_{ij}^{G_{Z}}$ for brevity.
\begin{equation}
\begin{aligned}
p^{G}_{ij}(\sigma _{i},\nu) &= g(d^G_{ij},\sigma _{i},\nu) \\
 &=C_{\nu}(1+\frac{d^G_{ij}}{\sigma _{i}\nu} )^{-\frac{(\nu+1)}{2} } 
\end{aligned}
\label{(8)}
\end{equation}
where the degree of freedom $\nu \in \mathbb{R}_+$, controls the heaviness of the tails; when it increases, the tails of the $t$-distribution become lighter, and smaller $\nu$ gives heavy tails so that even considerable outlier distances will have non-zero similarities. $C_{\nu}$ is a coefficient related to the gamma function $\Gamma(\cdot)$ about $\nu$.
\begin{equation}
C_{\nu}= \sqrt{2\pi} \frac{\Gamma (\frac{\nu+1}{2} )}{\sqrt{\nu\pi}\Gamma(\frac{\nu}{2} )}
\label{(9)}
\end{equation}
The parameter $\sigma_i$ in the $t$-distribution controls the scale of the distribution; lower $\sigma_i$ compresses and peaks the $t$-distribution, making the similarities drop off more rapidly with distance. 

After obtaining the graph node-to-node similarities, we can preserve the geometric information by designing a specific manifold learning loss. This manifold loss should penalize differences between similarities in a proportional way and account for imbalanced similarities. Therefore, a logistic manifold learning loss $L_R(P^{G_{Z',\mathcal{P}}},P^{G_{Z}})$ is adapted from:
\begin{equation}
L_{R}(a,b)=a\log{\frac{a}{b}+(1-a)\log{\frac{1-a}{1-b} } } \label{(11)}
\end{equation}
where $P^{G_{Z}}=\left \{ p_{ij}^{G_{Z}}|i,j=1,2, \cdots,n  \right \}$, similar for $P^{G_{Z',\mathcal{P}}}$.
Therefore, the final loss $L$ can be formalized as:
\begin{equation}
L=L_{task} + \beta L_R(P^{G_{Z',\mathcal{P}}},P^{G_{Z}})
\label{(12)}
\end{equation}
$\beta$ is a weight balance parameter, and $L_{task}$ is the task-related objective or loss function, such as cross-entropy loss. From Eq.~\ref{(12)}, we find that the manifold fold learning loss can be used in different protein-related tasks and even be extended into different deep neural networks if we want to preserve the graph topological features, like distance patterns, contacts, and connectivity, into latent embeddings. The structural constraints act as an inductive bias, regularizing the model to respect protein graph structure.

Considering the computational complexity of one message passing layer in this framework, it is $\mathcal{O}(n d_n )$, where $d_n$ is the average node degree, $d_n\le n$. As for the extra calculations for distance and similarity matrices that are introduced in the network based on deep manifold transformations, we only calculate the distances for neighbors of central nodes during the training process; thus, the complexity is $\mathcal{O}(n d_n )$. Considering the size of a dataset, using $B_s$ to denote the size of the batch, the final computational complexity is $\mathcal{O}(B_s n d_n )$.

\section{Experiments}
In this section, we extensively evaluate the generalization ability of the learned protein representations on a wide range of downstream applications, like protein fold classification, enzyme reaction classification, and protein-protein interaction identification, etc., where the related protein task and dataset settings are introduced in KeAP~\cite{Zhou2023ProteinRL}, GearNet~\cite{zhang2022protein} and CDConv~\cite{fancontinuous}. Our results are reported over five independent training runs. All codes are implemented using the PyTorch library and
run on an NVIDIA A100 GPU. The best results for each indicator
are shown in bold.

The hyperparameters related to the network are set the same across different datasets: Adam optimizer with learning rate of $0.001$, weight decay of $5 \times 10^{-4}$, epochs of $300$, Gaussian noise $\mu_k=0, \sigma_k=0.1$, it indicates trivial perturbation is introduced to the protein native structures. Radius threshold $r=10$, distance threshold $\Lambda=$20, as we deem that two residues are not in contact if the distance between them is larger than 20 Å. The other dataset-specific hyperparameters are determined by an AutoML toolkit NNI with the search spaces. For example, the loss weight hyperparameter is related to the value of the task-specific loss $\beta=\left \{ 1,0.1,0.01 \right \} $, the number of FC layers and message passing layers $L_F=\{0,1,2,3\}, L_M=\{2,4,6,8\}$. 
\subsection{Protein-Protein Interaction Identification}
The mean and standard deviation (std) values are reported for protein-protein interaction (PPI) identification, which are shown in Table~\ref{table2}. KeAP~\cite{Zhou2023ProteinRL} has not reported its std values; therefore, we show our and its reproduced results over five independent runs; other results are taken from SemiGNN-PPI~\cite{Zhao2023SemiGNNPPISM}. Following~\cite{Zhou2023ProteinRL}, we get the node embeddings from the pretrained model, KeAP, without the coordinates of structures. We can see that the proposed DMTPRL achieves the best or close to the best performance across these datasets. DMTPRL substantially outperforms prior methods, especially on SHS27K and SHS148K. Furthermore, both KeAp and DMTPRL use the exact pretrained node embeddings without protein structure information. Still, the proposed model can get higher averaged scores and lower std values on the prediction of PPI interactions, which demonstrates the effectiveness and convergence of the proposed deep manifold transformation loss.
\begin{table*}[tp]
  \caption{Comparisons on PPI identification with averaged (std) F1 scores are reported. $\left [ ^* \right ]$ denotes the results are taken from~\cite{Zhao2023SemiGNNPPISM} and $\left [ ^\dag \right ]$ means the reproduced results.}
  \centering
  \begin{tabular}{l|cc|cc|cc}
    \toprule
    \multirow{2}{*}{Methods} & \multicolumn{2}{c|}{SHS27K} & \multicolumn{2}{c|}{SHS148K} & \multicolumn{2}{c}{STRING} \\
    & BFS & DFS & BFS & DFS & BFS & DFS\\
    \midrule
    DNN-PPI~\cite{Hashemifar_Neyshabur_Khan_Xu_2018}$^*$ & 48.90(7.24) & 54.34(1.30) & 57.40(9.10) & 58.42(2.05) & 53.05(0.82) & 64.94(0.93)\\
    DPPI~\cite{Hashemifar_Neyshabur_Khan_Xu_2018}$^*$ & 41.43(0.56) & 46.12(3.02) & 52.12(8.70) & 52.03(1.18) & 56.68(1.04) & 66.82(0.29) \\
    PIPR~\cite{10.1093/bioinformatics/btz328}$^*$ & 44.48(0.75) & 57.80(3.24) & 61.83(10.23) & 63.98(0.76) & 55.65(1.60) & 67.45(0.34) \\
    GNN-PPI~\cite{Lv_Hu_Bi_Zhang_2021}$^*$ & 63.81(1.79) & 74.72(5.26) & 71.37(5.33) & 82.67(0.85) & 78.37(5.40) & 91.07(0.58) \\
    SemiGNN-PPI~\cite{Zhao2023SemiGNNPPISM} & 72.15(2.87) & 78.32(3.15) & 71.78(3.56) & \textbf{85.45}(1.17) & \textbf{80.84}(2.05) & 91.23(0.26) \\
    KeAP~\cite{Zhou2023ProteinRL}$^\dag$ & 75.74(5.14) & 79.39(3.52) & 73.19(7.13) & 82.54(2.10) & 78.41(4.79) & 88.54(1.78) \\
    DMTPRL (Ours) & \textbf{77.68}(2.19) & \textbf{79.44}(2.52) & \textbf{76.81}(4.83) & 83.34(2.13) & 78.51(1.83) & \textbf{90.26}(1.02) \\
    \bottomrule
  \end{tabular}
\label{table2}
\end{table*}

\begin{table}[htbp]
    \vspace{-1.0em}
  \caption{Accuracy ($\%$) on fold classification and enzyme reaction classification. $\left [ ^* \right ]$ denotes the results are taken from~\cite{fancontinuous}. }
  \centering
  \setlength{\tabcolsep}{0.5mm}{\begin{tabular}{l|ccc| c}
    \toprule
    \multirow{2}{*}{Method} & \multicolumn{3}{c|}{Fold Classification} & Enzyme \\
    \cmidrule(lr){2-4}
    &Fold & SuperFamily & Family & Reaction\\
    \midrule
     CNN~\cite{shanehsazzadeh2020transfer}$^*$ & 11.3 & 13.4 & 53.4 & 51.7 \\
     ResNet~\cite{RoshanRao2019EvaluatingPT}$^*$ & 10.1 & 7.21 & 23.5 & 24.1 \\ 
    LSTM~\cite{RoshanRao2019EvaluatingPT}$^*$ & 6.41 & 4.33 & 18.1 & 11.0 \\
    Transformer~\cite{RoshanRao2019EvaluatingPT}$^*$ & 9.22 & 8.81 & 40.4 & 26.6 \\
    GCN~\cite{kipf2016semi}$^*$ & 16.8 & 21.3 & 82.8 & 67.3 \\
    GAT~\cite{velickovic2017graph}$^*$ & 12.4 & 16.5 & 72.7 & 55.6 \\
    3DCNN$\_$MQA~\cite{derevyanko2018deep}$^*$ & 31.6 & 45.4 & 92.5 & 72.2 \\
     GVP~\cite{BowenJing2020LearningFP}$^*$ & 16.0 & 22.5& 83.8 & 65.5 \\
     ProNet-All-Atom~\cite{wanglearning} & 52.1 & 69.0 & 99.0 & 85.6 \\
     GearNet~\cite{zhang2022protein} & 28.4 & 42.6 & 95.3 & 79.4 \\
     GearNet-Edge~\cite{zhang2022protein} & 44.0 & 66.7 & 99.1 & 86.6 \\
     CDConv~\cite{fancontinuous} & 56.7 & 77.7 & \textbf{99.6} & 88.5 \\
     DMTPRL (Ours) & \textbf{59.9} & \textbf{78.7} & \textbf{99.6} & \textbf{89.3} \\
    \bottomrule
    \end{tabular}
    }
\vspace{-1.0em}
\label{table1}
\end{table}
\subsection{Protein Fold and Enzyme Reaction Classification}
Table~\ref{table1} compares the effects of different protein representation learning methods on the protein fold and enzyme reaction classification. DMTPRL achieves the best performance on these tasks compared with these currently effective methods. The biggest difference is that DMTPRL adds a manifold learning loss to constrain original embeddings in latent spaces to preserve distance relationships. These illustrate the effectiveness of the idea of graph-based topological structure preservation. 

\subsection{Amino Acid Contact Prediction}
This task is to predict whether two protein amino acids are in contact or not, given an input protein sequence. We do not use protein coordinate information in this task to prevent information leakage. We also get the node embeddings from~\cite{Zhou2023ProteinRL}, like in the PPI identification task. In Table~\ref{table3}, the sequence length of two selected amino acids is over 24; the proposed model can get higher contact predictions based on the pretrained protein embeddings. We believe the performance gains brought by DMTPRL can be attributed to the proposed deep manifold transformations by introducing an extra loss.
\begin{table}[htbp]
    \vspace{-1.0em}
	\centering
	\caption{Comparisons on amino acid contact prediction. $\left [ ^* \right ]$ denotes the results are taken from~\cite{Zhou2023ProteinRL}. P@L, P@L/2, P@L/5 denote the precision scores calculated upon top L, top L/2, and top L/5 predictions. }
    \setlength{\tabcolsep}{0.5mm}{\begin{tabular}{l|ccc}
    \toprule
    Methods& P@L & P@L/2 & P@L/5\\
    \midrule  
    ProtBert~\cite{AhmedElnaggar2021ProtTransTC}$^*$& 0.20 & 0.26 & 0.34 \\
    ESM-1b~\cite{AlexanderRives2019BiologicalSA}$^*$& 0.26 & 0.34 & 0.45 \\
    KeAP~\cite{Zhou2023ProteinRL} & 0.28 & \textbf{0.35} & 0.43 \\
    DMTPRL (Ours)& \textbf{0.30} & 0.34 & \textbf{0.45} \\

    \bottomrule 
    \end{tabular}
    }
    \label{table3}
    \vspace{-1.0em}
\end{table}

\section{Conclusion}
In this paper, we develop a simple yet effective framework by introducing manifold learning techniques to learn protein representations. Data augmentations and manifold transformations are applied to relieve the domain shift and data scarcity problems. The proposed deep manifold learning loss can constrain the latent embeddings. Experiments on standard benchmarks demonstrate our solution. In the future, we plan to pretrain on large-scale protein datasets by developing deep manifold transformations. 

\section{ACKNOWLEDGMENTS}
This work was supported by the National Key R\&D Program of China (No. 2022ZD0115100), the National Natural Science Foundation of China Project (No. U21A20427), and Project (No. WU2022A009) from the Center of Synthetic Biology and Integrated Bioengineering of Westlake University.

\vfill\pagebreak

\bibliographystyle{IEEEbib}
\bibliography{strings,refs}

\end{document}